\documentclass[aps,prd,twocolumn,nofootinbib,superscriptaddress]{revtex4}
\usepackage{amssymb,graphics,graphicx, epsfig}

\newcommand{\newc}{\newcommand}
\newc{\beqa}{\begin{eqnarray}}
\newc{\eeqa}{\end{eqnarray}}
\newc{\beq}{\begin{equation}}
\newc{\eeq}{\end{equation}}

\newc{\nonr}{\nonumber}
\newc{\ra}{\rightarrow}
\newc{\tri}{\triangle}
\newc{\PD}{\partial}
\newc{\lag}{{\cal{L}}}
\newc{\LH}{\hat{L}}
\newc{\RH}{\hat{R}}

\begin{document}

\title{A Very Narrow Shadow Extra Z-boson at Colliders}
\author{We-Fu Chang}
\email{wfchang@phys.sinica.edu.tw}
\affiliation{Institute of Physics, Academia Sinica, Nankang, Taipei, Taiwan 115}
\author{John N. Ng}
\email{misery@triumf.ca}
\author{Jackson M. S. Wu}
\email{jwu@triumf.ca}
\affiliation{Theory group, TRIUMF, 4004 Wesbrook Mall, Vancouver, B.C., Canada}

\date{\today}

\begin{abstract}
We consider the phenomenological consequences of a hidden Higgs sector extending the Standard Model (SM), in which the ``shadow Higgs'' are uncharged under the SM gauge groups. We consider a simple
$U(1)$ model with one Higgs singlet. One mechanism which sheds light on the shadow sector is the mixing between the neutral gauge boson of the SM and the additional $U(1)$ gauge group. The mixing happens through the usual mass-mixing and also kinetic-mixing, and is the only way the
``shadow $Z$'' couples to the SM. We study in detail
modifications to the electroweak precision tests (EWPTs) that the presence of such a shadow sector would bring,
which in turn provide constraints on the kinetic-mixing parameter, $s_\epsilon$,
left free in our model. The shadow $Z$ production rate at the LHC and ILC depends on $s_\epsilon$.
We find that observable event rate at both facilities is possible for a reasonable range of $s_\epsilon$ allowed by EWPTs.
\end{abstract}

\pacs{}

\maketitle

\section{Introduction}

In the pursuit of physics beyond the Standard Model (SM) it is very common to encounter one or
more Abelian gauge symmetry than the SM $U(1)$ hypercharge. Two familiar examples are the grand
Unified theories (GUTs) based on $SO(10)$ that breaks to $G_{SM}\times U(1)$,
where $G_{SM}$ is the SM gauge group and $E(6)$ which ultimately breaks to
$G_{SM}\times U(1)\times U(1)$. Because of their GUTs parentage the extra $Z$
 bosons from the breaking of the $U(1)$ symmetries have tree level couplings to
 the SM particles; in particular the fermions. This makes them highly visible
 and their phenomenology has been well studied~\cite{ExtraZ}. More recently extra
 dimensional models with extra $U(1)$ gauge symmetries in the brane world scenario
 are increasingly popular. A feature of this newer construction is that the extra
 $U(1)$ factors can be hidden from the visible sector. Hidden sectors are motivated
 also by studies in supersymmetry breaking mechanisms. Independent of the theoretical
 motivation, extra $Z$ bosons from hidden sector typically do not have direct couplings
 to the SM particles. Their phenomenology can be very different from visible extra $Z$'s.
  They are also harder to produce. As the start up of the LHC draws near, the search for
   extra $Z$ bosons is a high priority item due to their relatively clean signatures
   from Drell-Yan processes. Clearly it is important to include hard to find extra $Z$
    bosons in this search. Although these bosons have no direct couplings to SM particles,
  they can manifest themselves through mixings with the SM $Z$ boson, and so are not
 completely invisible. Since the mixing is crucial for phenomenology we construct the
simplest model of this kind to capture the physics of such an extra $Z$ boson.
 It has the gauge symmetry $G_{SM}\times U(1)_s$ where the subscript denotes ``shadow'';
 the name will become clear later. The SM fermions are singlet under $U(1)_s$. This $U(1)_s$ is broken by a shadow Higgs sector which is just the Abelian Higgs model with a complex scalar $\phi_s$. The $\phi_s$ field is a sinlget under $G_{SM}$ but interacts with the SM Higgs bosons via renormalizable interactions. The complete Lagrangian is given by
\beqa
\lag &=& \lag_{SM} - \frac{1}{4}X^{\mu\nu}X_{\mu\nu} - \frac{\epsilon}{2}B^{\mu\nu}X_{\mu\nu}\nonr\\
&+& \left|\left( \PD_{\mu}-\frac{1}{2}g_s X_{\mu}\right) \phi_s\right|^2 - V(\phi_s,\Phi) \,,
\eeqa
where $B_{\mu\nu}$ is the field strength tensor of the SM hypercharge $U(1)_Y$, $\Phi$ is the SM Higgs
field, and $g_s$ is the gauge coupling of the shadow $U(1)_s$. For simplicity we have normalized the shadow charge of $\phi_s$ to unity. The kinematic mixing of the two $U(1)$'s is parameterized by
$\epsilon$, which a priori need not be a small number. For a visible extra $Z$ this mixing term is expected to be only induced at the loop level~\cite{Holdom}, and thus $|\epsilon|\ll 1$ is generally assumed in its phenomenological studies~\cite{ZZpMix}. However, this need not be the case here. Indeed, a calculation in string theory of the mixing-generating vacuum polarization diagram shows that in general, one can expect kinetic mixing effects on the order of $10^{-4}\,\sim\, 10^{-2}$ at the weak scale (barring accidental cancellations in the tree level spectrum)~\cite{string}.
Given the theoretical significance outlined above, we shall leave $\epsilon$ as a free parameter to be constrained by experiments in particular the electroweak precision tests (EWPTs). Now it is well known that the kinetic terms including the mixing can be recast into canonical form through a $GL(2)$ transformation. Explicitly, this is given by
\beq
\left(\begin{array}{c} X\\ B \end{array}\right)
= \left(\begin{array}{cc} c_\epsilon & 0 \\ - s_\epsilon & 1 \end{array}\right)
\left(\begin{array}{c} X'\\ B' \end{array}\right) \,,
\eeq
where
\beq\label{Eq:KinMix}
s_\epsilon = \frac{\epsilon}{\sqrt{1 - \epsilon^2}}\,,\;
c_\epsilon = \frac{1}{\sqrt{1 - \epsilon^2}} \,.
\eeq
After spontaneous symmetry breaking (SSB) $X'$ and $B'$ will mix resulting in a shift in the SM $Z$ mass. The physical bosons are now linear combinations of the two. The photon will remain massless, and the $W$ bosons will be unchanged from the SM. This is expected since the shadow sector only interacts with $U(1)_Y$ through the shadow Higgs interactions. The details of this symmetry breaking is given in Sec.~\ref{sec:SBShW}. Feynman rules for the model are also given there. In this paper we focus on the phenomenology of the physical shadow sector neutral boson, $Z_s$. Since we are interested in collider physics we shall assume that vacuum expectation value (VEV) of $\phi_s$ is of the order of the weak scale or higher. In Sec.~\ref{sec:Phem} we study the impact $Z_s$ has on electroweak precision measurements. From these, as well as anomalous magnetic moment of the muon and recent results from M\o ller scattering, we derive constraints on the parameters of our model, in particular on $s_\epsilon$. We employ a conservative strategy and demand that the fits to the data are not much worse than that of the SM. With these limits in hand we explore the prospect of observing the $Z_s$ at the LHC and the ILC in Sec.~\ref{sec:LHCILC}. Finally we give our conclusions in Sec.~\ref{sec:Concl}. Recent work with an extra $Z$ similar to ours is given in~\cite{LHCprobe} and the older literature can be found in~\cite{LowEE6,PhemZp}.

\section{Symmetry Breaking and The Shadow World}\label{sec:SBShW}

The most general renormalizable $G_{SM}\times U(1)_s$ invariant scalar potential is:
\beqa
V(\Phi, \phi_p) &=&  \mu_s^2 \phi_s^*\phi_s +\lambda_s (\phi_s^* \phi_s)^2
+2 \kappa \left(\Phi^\dag \Phi\right) \left(\phi_s^*\phi_s\right)\nonr\\
&& +\mu^2\Phi^\dag\Phi+\lambda (\Phi^\dag \Phi)^2 \,.
\eeqa
This Higgs potential is also used in phantom Higgs models~\cite{HiddenH}. After SSB the scalars
acquire nonzero VEV,
\beq
\left\langle \Phi\right\rangle =\frac{1}{\sqrt{2}}\left( \begin{array}{c}
  0 \\v_0 \end{array}\right)\,,\;
\left\langle \phi_p \right\rangle = \frac{v_s}{ \sqrt{2}} \,,
\eeq
with
\beq
v_0^2 = -\frac{\lambda_s \mu^2 -\kappa \mu_s^2}{\lambda\lambda_s - \kappa^2}\,,\;
v_s^2 = -\frac{\lambda \mu_s^2 -\kappa \mu^2}{\lambda\lambda_s - \kappa^2} \,.
\eeq
To ensure that the potential is bounded from below and the above values correspond to a global minimum we require $\lambda,\lambda_s > 0$ and $\kappa>0$.

The $SU(2)_L \times U(1)_Y \times U(1)_s$ symmetry is broken down to $U(1)_{QED}$. This pattern of breaking is peculiar in that the mass of the $W$ boson remains as in the SM, i.e.
$M_W = (g_2 v_0 )/2$. In the neutral sector we have a massless photon and two massive neutral bosons which are not yet in the mass eigenbasis. The usual SM definition: $\tan\theta_W = g_Y/g_2$,
electric charge $e = g_2\sin\theta_W$, and $Q^f_{L,R} = T^3_{L,R}+Y^f_{L,R}$ remain intact.

For the neutral gauge bosons the transformation between the weak and mass basis is given by the following rotation:
\beq
\left( \begin{array}{c} B'\\A_3\\ X' \end{array}\right)
=\left( \begin{array}{ccc} c_W & -s_W& 0\\
s_W & c_W & 0\\ 0&0&1 \end{array}\right)
\left( \begin{array}{ccc} 1&0&0\\ 0& c_\eta & -s_\eta\\
0& s_\eta & c_\eta \end{array}\right)
\left( \begin{array}{c} \gamma \\ Z \\ Z_s \end{array}\right) \,,
\eeq
where $s_W$ ($c_W$) denotes $\sin\theta_W$ ($\cos\theta_W$) and similarly for the rotation angle
$\eta$. The first rotation is the standard one that gives rise to the SM $Z$ and the second one diagonalizes the mixing of the two $Z$ bosons. The mixing angle is given by
\beq
\tan 2\eta = \frac{2 s_W s_\epsilon}{c_W^2 (M_3/M_W)^2 + s_W^2 s_\epsilon^2 - 1} \,,
\eeq
where $M_3 \equiv (g_sv_s)/2$. For small $s_\epsilon$ and $c_W M_3 > M_W$, $\eta < \epsilon$.
The masses for the two massive neutral gauge bosons are readily found to be
\beqa
\label{Zmasses}
M^2_{Z,Z_{s}}= {M_W^2 \over 2 c_W^2}
\left\{ \left({c_W M_3 \over M_W}\right)^2 + 1 + s_W^2
s_\epsilon^2\;\;\;\right.\nonr\\
\left.\mp \sqrt{\left[(c_W M_3/M_W)^2 - 1+ s_W^2 s_\epsilon^2\right]^2 + 4 s_W^2 s_\epsilon^2}
\;\right\} \,,
\eeqa
For the case where $M_3 > M_W$ the $Z$-$Z_s$ mixing is proportional to $s_\epsilon$, which is related to the amplitude of the kinetic mixing term $B^{\mu\nu}X_{\mu\nu}$.

The most stringent constraints on any extra $Z$ model come from EWPTs, and so we consider next the gauge
fermion couplings. These can be readily read off from the Lagrangian. For the photon ($A^\mu$), the
SM result is retained as it should:
\beq
A^\mu \bar{f}f\; : \; i \gamma^\mu e Q_f \,.
\eeq
For $Z$, $Z_s$, the coupling are slightly different from the SM, but still flavor
universal:
\beqa
\label{eq:zff}
Z^\mu \bar{f}f\; : \; i \gamma^\mu \frac{ g_2}{c_W}
\left[ \left( c_\eta  g^L_f   - s_\eta s_W s_\epsilon  Y_f^L  \right)
\LH\;\right.\nonr\\
\left. +\left( c_\eta g^R_f- s_\eta s_W s_\epsilon  Y_f^R   \right) \RH \right] \,,\\
\label{eq:zprimeff}
Z_s^\mu \bar{f}f\; : \;i \gamma^\mu \frac{ g_2}{c_W}
\left[ \left( -s_\eta  g^L_f   - c_\eta s_W s_\epsilon  Y_f^L  \right)
\LH\;\right.\nonr\\
 \left. +\left( -s_\eta g^R_f- c_\eta s_W s_\epsilon  Y_f^R   \right) \RH \right] \,,
\eeqa
where $g_{L,R}^f = T^3(f_{L,R}) - s_W^2 Q^f$ is the coupling of the SM $Z$ to fermions. We see that the neutral current couplings are not only rotated as indicated by the $c_\eta$ factor, but also contain an extra piece proportional to the fermion hypercharge due to $U(1)$-$U(1)_s$ mixing. Hence we need to reexamine the electroweak precision data using the full couplings as well as taking into account the effects due to virtual $Z_s$ exchanges.
\begin{figure}[htbp]
\centering
\includegraphics[width=0.35\textwidth]{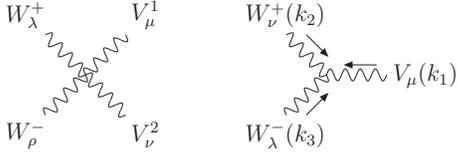}
\caption{The triple and quartic gauge vertices and the momentum
labelling.} \label{fig:34gauge_fig}
\end{figure}
For the pure gauge sector, it's straightforward to work out the Feynman rules. For example, the Feynman rules for 4 gauge boson $(V^1_\mu V^2_\nu W^+_\lambda W^-_\rho) $ vertex read:
\beq
-i g_2^2  C_4^{V_1,V_2} \left[ 2 g_{\mu\nu}g_{\lambda\rho}
-g_{\mu\lambda}g_{\nu\rho} -g_{\mu\rho} g_{\nu\lambda} \right] \,,
\eeq
where the factor $C_4$ are listed below:
\begin{center}
\begin{tabular}{|c|c|c|c|c|c|}
\hline
$(\gamma,\gamma)$ & $(Z,Z)$ & $(Z_s,Z_s)$ & $(\gamma, Z)$ & $(\gamma,Z_s)$ & $(Z,Z_s)$\\
\hline
$s_W^2$ & $c_W^2 c_\eta^2$ & $c_W^2 s_\eta^2$ & $s_W c_W c_\eta$ & $-s_W c_W s_\eta$ &
$-c_W^2 s_\eta c_\eta $\\
\hline
\end{tabular}
\end{center}
Similarly, the Feynman rules for triple gauge coupling for $V_\mu(k_1)W^+_\nu(k_2)W^-_\lambda (k_3)$,
where all momentum $k$'s go into the vertex, read:
\beq
-i g_2 C_3^V \left[(k_1-k_2)_\lambda g_{\mu\nu}
+(k_2-k_3)_\mu g_{\nu\lambda}
+(k_3-k_1)_\nu g_{\lambda\mu} \right] \,,
\eeq
and for different $V$s the $C_3$'s are
\begin{center}
\begin{tabular}{|c|c|c|}
\hline
$\gamma$ & $Z$ & $Z_s$\\
\hline
$ s_W $ & $ c_W c_\eta $ & $-c_W s_\eta$\\
\hline
\end{tabular}
\end{center}

Now a few remarks about the scalar sector. The SM Higgs doublet has 4 degrees of freedom (DOF) and the shadow scalar has 2 DOF. After SSB in both sectors, one DOF of each scalar becomes massive physical scalar. So we are left with $3+1$ massless DOFs which will be eaten by two $W$, one $Z$, and one $Z_s$, and the DOF budget is balanced. Therefore, the shadow world gives us one extra neutral scalar and no charged scalars. How heavy they are is an interesting question. We will assume here the lighter one is SM-like and has a mass greater than 114 GeV and the heavier one
is more than 200 GeV. This amounts to assuming $v_s\gtrsim v_0$ and no fine tuning of the Higgs parameters. In the basis of $\{h^0_{SM}, h^0_s \}$, the mass matrix for these two neutral scalars is
\beq
\left(\begin{array}{cc} \lambda  v_0^2 & \kappa v_0 v_s\\
\kappa v_0 v_s & \lambda_p v_s^2   \end{array}\right) \,.
\eeq
It can be diagonalized by a rotation
\beq
\left( \begin{array}{c} h^0_{SM}\\ h^0_s \end{array}\right)
= \left( \begin{array}{cc} \cos\alpha & \sin\alpha\\
-\sin\alpha& \cos\alpha \end{array}\right)
\left( \begin{array}{c} h^0_1\\ h^0_2 \end{array}\right) \,,
\eeq
and the mixing angle satisfies
\beq\label{Eq:HiggsMix}
\tan (2\alpha) = \frac{2 \kappa v_0 v_s}{\lambda_s v_s^2 - \lambda v_0^2} \,,
\eeq
with mass square
\beq
m^2_{1,2}=\frac12\left[\lambda_s v_s^2 + \lambda v_0^2 \pm
\sqrt{(\lambda_p v_p^2- \lambda v_0^2)^2 + 4\kappa^2 v_s^2
v_0^2}\;\right] \,.
\eeq

The Feynman rules for the scalar sector can be readily worked out. For instance, in the mass basis of the fermions, the scalar-fermion couplings are given by
\beqa
h_1^0 \bar{f} f &:& -i \cos\alpha \frac{g_2 m_f}{\sqrt{2} M_W}\,,\\
h_2^0 \bar{f} f &:& -i \sin\alpha \frac{g_2 m_f}{\sqrt{2} M_W}\,,
\eeqa
and for the gauge-scalar couplings one has
\beqa
h_1^0 Z^\mu Z^\nu &:& \frac{i g_2 M_W}{c_W^2}c_\alpha
(c_\eta +s_W s_\epsilon s_\eta)^2 g^{\mu\nu}\,, \\
h_1^0 Z_s^\mu Z_s^\nu &:& \frac{i g_2 M_W}{c_W^2}c_\alpha
(s_\eta - s_W s_\epsilon c_\eta)^2 g^{\mu\nu}\,,\\
h_1^0 Z_s^\mu Z^\nu &:& -\frac{i g_2 M_W}{c_W^2}c_\alpha g^{\mu\nu}\nonr\\
&& \times(c_\eta + s_W s_\epsilon s_\eta)(s_\eta - s_W s_\epsilon c_\eta)\,,\\
h_1^0 W^{+,\mu} W^{-,\nu} &:& i g_2 M_W c_\alpha g^{\mu\nu}\,.
\eeqa
For $h_2^0$, the ``shadow Higgs'', $c_\alpha$ is replaced by $s_\alpha$ in the above expression.

\section{Phenomenology}\label{sec:Phem}

We now perform a systematic phenomenological study using the Feynman rules derived previously. We present the analytical result involving parameters of the shadow sector; numerical values are summarized in Table~\ref{tab:EWPT_tab} at the end of the section. We begin with the anomalous magnetic moment of the muon.

\subsection{Muon $g-2$}
 The one-loop $Z$ boson contribution to the anomalous magnetic moment of a charged lepton is now
 different from the SM due to modification of its coupling to fermions.
 Furthermore, the same  one-loop diagram with $Z_s$ running in the loop contributes as well.
 Plugging in the gauge-fermion interactions obtained in the previous section,
 the SM one-loop $Z$ boson contribution now shifts by an amount:
\beqa
 \delta a_\mu &=& s_\eta^2 \tri a_Z^{SM}\left[1 -
\left(\frac{M_Z}{M_{Z_s}}\right)^2\right]\\
&+& \frac{g_2^2 m_\mu^2}{8\pi^2 M_Z^2 c_W^2}
\left[ \left(s_W^2 - \frac13\right) s_\eta c_\eta s_W s_\epsilon
+\frac16 s_W^2 s_\epsilon^2 s_\eta^2 \right]\nonr \\
&-& \frac{g_2^2 m_\mu^2}{8\pi^2 M_{Z_s}^2 c_W^2}
\left[  \left(s_W^2-\frac13\right)s_\eta
c_\eta s_W s_\epsilon -\frac16 s_W^2 s_\epsilon^2
s_\eta^2\right]\nonr\\
&=& 1.94\times 10^{-9}
\left(1-\frac{M_Z^2}{M_{Z_s}^2}\right)\nonr\\
&& \times[s_\eta^2 -0.2354 s_\eta c_\eta s_\epsilon +0.1850 s_\epsilon^2
s_\eta^2] \,,
\eeqa
where we have set $s_W^2=0.2311$.
The $W$ boson and other SM contributions remain the same.
The two Higgs bosons and higher loop diagrams contribute negligibly.
We shall see later that $\delta a_\mu$ above does not give better constraints than EWPT.

\subsection{NuTeV}
The NuTeV experiment measures the ratio of neutral current to charged current cross-sections
in deep-inelastic $\nu_\mu$-nucleon scattering~\cite{NuTeVexpt}. As was suggested by
Paschos-Wolfenstein~\cite{PWobs} to reduce theoretical and systematic uncertainties, the
precision observable to measure  at NuTeV is
\beqa
R_{PW}
&=& \frac{\sigma(\nu_\mu N \rightarrow \nu_\mu X) -
               \sigma(\bar{\nu}_\mu N \rightarrow \bar{\nu}_\mu X)}
              {\sigma(\nu_\mu N \rightarrow \mu X) -
               \sigma(\bar{\nu}_\mu N \rightarrow \mu^+ X)}\nonr\\
&=& (g^N_L)^2 - (g^N_R)^2 = \frac{1}{2} - s_W^2 \,,
\eeqa
where $(g^N_{L,R})^2 = (g^u_{L,R})^2 + (g^d_{L,R})^2$. 
Since the shadow world affects only the neutral current processes, presence of a new neutral
gauge boson will only affect the numerator of $R_{PW}$. For the $\nu_\mu q$ elastic scattering
process, the squared amplitude receives corrections from the modifications in the couplings of
the SM $Z$, and from contributions due to the exchange of the virtual $Z_s$.
Incorporating these effects due to the $Z_s$, a straightforward calculation shows that the
numerator of $R_{PW}$ is proportional to
\beqa\label{Eq:RPWnum}
\sum_{q=u,d}\left\{
\big[(g_L^q)^2 - (g_R^q)^2\big] +
2\frac{M_Z^2}{M_{Z_s}^2} g_L^\nu \big[x_L^q g_L^q - x_R^q g_R^q\big]\right\}\nonr\\
 + \mathcal{O}\left({M_Z^4\over M_{Z_s}^4}\right)\,,
\eeqa
where $x^f_{L,R}=-(s_\eta g^f_{L,R}+c_\eta Y^f s_W s_\epsilon)$ is the $Z_{s}f\bar{f}$
coupling given in Eq.(\ref{eq:zprimeff}),
and $q^2 \ll M_Z^2,M_{Z_s}^2$ is assumed, with $q$ the momentum transfer. The first term in
Eq.~(\ref{Eq:RPWnum}) is the SM result, while the second term comes from the SM-shadow
interference; we have ignored the term suppressed by $(M_Z/M_{Z_s})^4$. Note that for the
isoscalar targets considered at NuTeV, the sum is over $u$ and $d$ quark distributions. Assuming
that $M_{Z_{s}}\gg M_Z$, only the SM contribution need be kept while taking into account the
modifications to the Z-fermion coupling. In terms of the mixing parameters of our model, the
effective nucleon coupling to the SM $Z$ is given by:
\beqa
(g^N_L)^2 \simeq (g^N_L)^2_{SM}\left[ 1-2s_\eta^2 +1.0028 s_\eta c_\eta s_\epsilon
\right]\;,\nonr\\
(g^N_R)^2\simeq (g^N_R)^2_{SM}\left[ 1-2s_\eta^2 +5.1218  s_\eta c_\eta s_\epsilon
\right]\;.
\eeqa
Note that the coupling to neutrinos is absorbed into the effective couplings here.

\subsection{M\o ller Scattering at SLAC}
The SLAC E158 M\o ller scattering experiment~\cite{Anthony:2005pm} measures the parity violating asymmetry,
\beq
A_{PV} = \frac{\sigma_L-\sigma_R}{\sigma_L+\sigma_R} \,,
\eeq
at momentum transfer $Q^2=0.026\,\textrm{GeV}^2$. The subscripts $L$ and $R$ denote the incident electron polarization. At tree level, the asymmetry is, up to ${\cal O}(g_2^2)$:
\beq
A_{PV} \simeq \frac{G_F s}{\sqrt{2}\pi \alpha}\frac{y(1-y)}{1+y^4+(1-y)^4}
(g_L^{e2}-g_R^{e2}) \,,
\eeq
where $y = Q^2/s \simeq 0.6$.

The denominator in the above expression represent the leading M\o ller cross-section due to photon exchange, and the numerator is the parity violation due to photon-Z interference. It is easy to extend it to include the $Z_s$ contribution. We only need to keep the photon-$Z_s$ interference term:
\beq
A_{PV}^{Z_s} \simeq \frac{G_F s}{\sqrt{2}\pi\alpha}\frac{y(1-y)}{1+y^4+(1-y)^4}
(x_L^{e2}-x_R^{e2})\left(\frac{M_Z}{M_{Z_s}}\right)^2
\eeq

Assuming that $M_{Z_s}\gg M_{Z}$, the $Z_s$ effect can be ignored. From the modified $Z e\bar{e}$ coupling we have
\beqa
\frac{\delta A_{PV}}{A_{PV}}
&\simeq& -s_\eta^2 - \frac34 \frac{s_\eta^2 s_W^2 s_\epsilon^2}{\frac14 -s_W^2}
- c_\eta s_\eta s_W s_\epsilon \frac{s_W^2 + \frac12}{\frac14 - s_W^2}\\
&\simeq & -[s_\eta^2 + 9.171 s_\eta^2 s_\epsilon^2 +18.60 c_\eta s_\eta s_\epsilon] \,.
\eeqa
This translates into
\beqa
\delta \sin^2 \theta_{eff}
&\simeq& -\frac{1 - 4 s_W^2}{4}\frac{\delta A_{PV}}{A_{PV}}\nonr\\
&\simeq& 0.019\,[s_\eta^2 +9.17 s_\eta^2 s_\epsilon^2 +18.60 c_\eta s_\eta s_\epsilon] \,.
\eeqa

\subsection{Atomic Parity Violation}
In the atomic system, the exchange of SM $Z$ boson will generate the parity violating $M1$ transition. This optical line can be accurately measured and used to compare with the theoretical prediction~\cite{QWexp}. Since the momentum transferred by the $Z$ boson is much smaller than nuclear mass, it can sense the weak charge of all the quarks coherently. The relevant quantity is:
\beqa
Q_W = -4(g^e_L-g^e_R)[(2Z+N)(g^u_L+g^u_R)\nonr\\
+(2N+Z)(g^d_L+g^d_R)]\,.
\eeqa
The expression for the contribution from an extra neutral gauge boson, $X$,
is same as above with $g_{L,R}$ changed to the corresponding couplings for $X$ and multiplied by an extra mass factor
$m_Z^2/m_X^2$.

If $Z_s$ is much heavier than the SM $Z$, its tree-level effect goes like
$s_\epsilon^2(M_Z/M_{Z_s})^2$, which can be again ignored. Therefore, the leading change to $Q_W$ comes from the modification to the couplings of SM $Z$ boson to fermions.
At tree level, we have for $Cs^{133}_{55}$
\beq
\frac{\delta Q_W}{Q_W} \simeq
-s_\eta^2 +0.7605 s_\eta^2 s_\epsilon^2 +2.0627 c_\eta s_\eta s_\epsilon \,,
\eeq
and for $Tl^{205}_{81}$
\beq
\frac{\delta Q_W}{Q_W} \simeq
-s_\eta^2 +0.7195 s_\eta^2 s_\epsilon^2 +1.9774 c_\eta s_\eta s_\epsilon \,.
\eeq

\subsection{Asymmetries at LEP}
Consider first the general expression of 
differential cross section for the process  $e^- +e^+\ra f^- +f^+$ mediated by more than one neutral gauge boson. If we ignored all the light fermion masses, the differential cross section for $f^-$ deflected from the incident $e^-$ direction by angle $\theta$ are given by:
\begin{widetext}
\beqa
{d \sigma\over d(\cos\theta)}= {N_c \pi \alpha^2 \over 8
s}\sum_{X,Y}K_{XY}
\left\{ (1+\cos\theta)^2\left(g^e_{XL} g^e_{YL}g^f_{XL} g^f_{YL}+g^e_{XR} g^e_{YR}g^f_{XR}
g^f_{YR}\right)\right.\nonr\\
\left.+(1-\cos\theta)^2\left(g^e_{XL} g^e_{YL}g^f_{XR} g^f_{YR}+g^e_{XR} g^e_{YR}g^f_{XL} g^f_{YL}\right)\right\} \,,
\eeqa
where indices $X,Y$ run over all neutral gauge bosons and the $K$'s are kinematic factors. For instance, $K_{\gamma\gamma}= Q_f^2$, and for $X,Y\neq \gamma$,
\beqa
K_{X\gamma}&=& \left(\frac{-2 Q_f}{c_W^2 s_W^2}\right)
{s(s-M_X^2)\over[ (s-M_X^2)^2 +M_X^2 \Gamma_X^2]} \,,\\
K_{XX}&=& \left({1\over c_W^4 s_W^4 }\right){s^2\over (s-M_X^2)^2 +M_X^2 \Gamma_X^2} \,,\\
K_{XY}&=& \left({2\over c_W^4 s_W^4 }\right)
{ s^2[(s-M_Y^2)(s-M_X^2)+ M_Y \Gamma_Y M_X \Gamma_X ]\over
[ (s-M_Y^2)^2 +M_Y^2 \Gamma_Y^2 ][(s-M_X^2)^2 +M_X^2 \Gamma_X^2]} \,,
\eeqa
where $M$ and $\Gamma$ are the mass and width of the neutral gauge boson respectively. For photon coupling, it has been normalized to be $g_L=g_R=1$. For other neutral gauge boson coupling, the coupling is normalized by the SM  strength $(g_2/c_W)$. The forward-backward asymmetry is then given by
\beq
A^f_{FB} =\frac34\;
\frac{\sum_{X,Y} K_{XY}(g^e_{XL} g^e_{YL} - g^e_{XR} g^e_{YR})
\left(g^f_{XL} g^f_{YL} - g^f_{XR} g^f_{YR}\right)}
{\sum_{X,Y} K_{XY}(g^e_{XL} g^e_{YL} + g^e_{XR} g^e_{YR})
\left(g^f_{XL} g^f_{YL} + g^f_{XR} g^f_{YR}\right)} \,.
\eeq

In SM, the fermion's left-right asymmetry can be derived from the left-right forward-backward asymmetry:
\beq
A_f = \frac43\; { \sigma^f_{LF}-\sigma^f_{LB} - \sigma^f_{RF}+ \sigma^f_{RB}
\over  \sigma^f_{LF}+\sigma^f_{LB} + \sigma^f_{RF}+ \sigma^f_{RB}} \,,
\eeq
where $L$ ($R$) stands for the left(right)-handed incident electron, and $F$ ($B$) stands for the forward (backward) direction, $\cos\theta>0\,(<0)$, of the final state fermion. When more than one massive neutral gauge bosons are present, the effective left-right asymmetry becomes:
\beq
A_f =
{\sum_{X,Y} K_{XY}(g^e_{XL} g^e_{YL} + g^e_{XR} g^e_{YR} )
\left(g^f_{XL} g^f_{YL} - g^f_{XR} g^f_{YR} \right)
\over
\sum_{X,Y} K_{XY}(g^e_{XL} g^e_{YL} + g^e_{XR} g^e_{YR} )
\left(g^f_{XL} g^f_{YL} + g^f_{XR} g^f_{YR} \right)} \,.
\eeq
\end{widetext}

At the Z pole, $K_{ZZ}=42298.1$. The other $K$ factors are very small and all of them can be ignored except $K_{XX}$, which has a very narrow and high spike when $M_{Z_s} \sim M_Z$. However, it drops very quickly when $M_{Z_s}$ fall outside the Z width. Therefore, for a heavy $Z_s$ we only need to consider the modification of the SM $Zf\bar{f}$ coupling and its effect on the precision measurement.

In Table~\ref{tab:EWPT_tab}, we summarize the current LEP, NuTeV, and SLAC M\o ller status
(from~\cite{Eidelman:2004wy}) and our prediction. The second column $\tri_{exp}$ is the experimental fractional  deviation from the SM prediction and the combined theoretical and experimental uncertainty, $(\delta_{exp})$, is shown in the parenthesis. The third column gives the fractional deviation from the SM our shadow model predicts.
\begin{table}[htbp]
\begin{center}
\begin{tabular}{ccc}
  \hline\hline
  Quantity &  $\tri_{exp}\equiv\left({\mbox{Exp}\over \mbox{SM}}\right)-1$
  &  $\tri_s\equiv\left({\mbox{Model}\over \mbox{SM}}\right)-1 $\\
  \hline
   $\Gamma_Z$ & $-0.0008(10)$ &$-s_\eta^2 +0.5730 s_\eta c_\eta s_\epsilon $ \\
   $\sigma_{had}$ & $+0.0017(9)$ &$ +0.02593 s_\eta c_\eta s_\epsilon $ \\
   $\Gamma(had)$ & $+0.0005(13)$ & $-s_\eta^2 +0.4327 s_\eta c_\eta s_\epsilon $\\
   $\Gamma(inv)$ & $-0.0056(30)$ & $-s_\eta^2 +0.961 s_\eta c_\eta s_\epsilon $\\
   $\Gamma(l\bar{l})$ & $-0.00048(107)$ & $-s_\eta^2 +0.7393 s_\eta c_\eta s_\epsilon $ \\
   $R_e$ & $+0.0026(25)$&  $-0.3065 s_\eta c_\eta s_\epsilon$\\
   $R_\mu$ & $+0.0016(17)$&  $-0.3065 s_\eta c_\eta s_\epsilon$ \\
   $R_\tau$ & $-0.0013(23)$& $-0.3065 s_\eta c_\eta s_\epsilon$ \\
   $R_b$ & $+0.0034(31)$&  $+0.0676 s_\eta c_\eta s_\epsilon$\\
   $R_c$ & $-0.0019(174)$&  $-0.1306 s_\eta c_\eta s_\epsilon$\\
   $A^{(0,e)}_{FB}$ & $-0.108(154)$&  $-38.67 s_\eta c_\eta s_\epsilon$ \\
     $A^{(0,\mu)}_{FB}$ & $+0.039(82)$&  $-38.67 s_\eta c_\eta s_\epsilon$ \\
       $A^{(0,\tau)}_{FB}$ & $+0.156(106)$&  $-38.67 s_\eta c_\eta s_\epsilon$ \\
   $A^{(0,b)}_{FB}$ & $-0.034(17)$&  $-19.59 s_\eta c_\eta s_\epsilon$\\
   $A^{(0,c)}_{FB}$ & $-0.043(48)$& $-21.24 s_\eta c_\eta s_\epsilon$\\
   $A^{(0,s)}_{FB}$ & $-0.055(111)$& $-19.59 s_\eta c_\eta s_\epsilon$\\
   $A_e$ & $+0.028(16)$&  $-19.335 s_\eta c_\eta s_\epsilon$\\
     & $+0.049(42)$&  $-19.335 s_\eta c_\eta s_\epsilon$\\
     & $+0.018(34)$&  $-19.335 s_\eta c_\eta s_\epsilon$\\
      $A_\mu$ & $-0.035(102)$&  $-19.335 s_\eta c_\eta s_\epsilon$\\
       $A_\tau$ & $-0.076(102)$&  $-19.335 s_\eta c_\eta s_\epsilon$\\
              & $-0.022(30)$&  $-19.335 s_\eta c_\eta s_\epsilon$\\
     $A_b$ & $-0.010(21)$&  $-0.251 s_\eta c_\eta s_\epsilon$\\
       $A_c$ & $+0.003(39)$&  $-1.909 s_\eta c_\eta s_\epsilon$\\
         $A_s$ & $-0.043(97)$&  $-0.251 s_\eta c_\eta s_\epsilon$\\
         \hline
NuTeV\ \ \ \\
    $(g_L^N)^2 $ & $-0.013(5)$ & $-2s_\eta^2 +1.0028 s_\eta c_\eta s_\epsilon$ \\
   $(g_R^N)^2 $ & $+0.023(36)$ & $-2s_\eta^2 +5.1218  s_\eta c_\eta s_\epsilon$\\
  \hline
SLAC M\o ller \\
 $ \sin^2\theta_{eff}$ & $+0.007(22)$
 & $0.019 s_\eta^2 + 0.353s_\eta c_\eta
 s_\epsilon$\\
 \hline
$Q_W(Cs^{133})$ & $-0.0068(66)$& $-s_\eta^2  +2.0627 c_\eta s_\eta s_\epsilon$\\
$Q_W(Tl^{205})$& $-0.0018(317)$&$-s_\eta^2 +1.9774 c_\eta s_\eta s_\epsilon$\\
   \hline\hline
   \end{tabular}
\end{center}
\caption{The comparison of experimental values and theoretical prediction, up to
${\cal O}(s_\epsilon^2)$, for various EWPT observables.}
\label{tab:EWPT_tab}
\end{table}

To give a measure of how well our model fits the data from EWPTs compare to that using purely the SM,
we define a number, $\chi_2$, that measures the deviation between theory and experiment,
\beq\label{chi2}
\chi_2(s_\epsilon, M_3) \equiv \sum_i \left( { \tri_{exp}^i- \tri_s^i\over
\delta^i_{exp}}\right)^2\;.
\eeq
For SM only, the deviation is
\beq
\chi_2^{SM} = \chi_2(0,M_3) = \chi_2(s_\epsilon,\infty) = 34.908 \,.
\eeq
The parameter space allowed for $s_\epsilon$ and $M_3$ (or $s_\eta$) will be determined by doing a simple least square fit. We are interested in getting a solution which can lower the $\chi_2$, or
\beq
\label{delchi2}
\tri \chi_2 \equiv \chi_2(s_\epsilon,M_3)-\chi_2^{SM}\leq 0 \,,
\eeq
indicating a better fit than the SM. However, for $M_3 > 200$ GeV, our numerical search did not find any parameter space which can improve the global fitting listed in Table~\ref{tab:EWPT_tab}. This is easy to understand. For $M_{Z_s}\gg M_Z$, $s_\eta \ll s_\epsilon$ and the $s_\eta^2$ corrections in the third column of the table is not important. One sees that about half of the EW observables get the
wrong sign corrections. Therefore at most we can only make gain and loss balanced.

\begin{figure}[ht]
\centering
\includegraphics[width=0.35\textwidth]{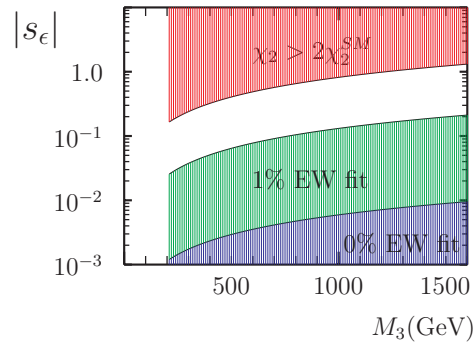}
\caption{The bound on $s_\epsilon$ and $M_3$ from EWPT.
The upper band is the excluding region by too large a deviation from SM,
$\chi_2> 2 \chi_2^{SM}$.
The lower band in the parameter space  gives comparable to SM results in the global fit.
And the middle one is the allowed region where $(\tri\chi_2/\chi_2^{SM})<0.01$.}
\label{fig:GFit_BM3_fig}
\end{figure}

Therefore, we demand that the allowed parameter space does not make the fitting worse and this is shown as the lower band in Fig.~\ref{fig:GFit_BM3_fig}.
The allowed parameter region is approximately given by
\beq\label{M3limit}
M_3 > 166 |s_\epsilon|\,\mbox{TeV}\quad\mbox{or}\quad
|s_\epsilon| < 0.006 \left({M_3 \over 1 \mbox{TeV}} \right)\,.
\eeq
However, if we relax the global fitting a little,
$(\tri \chi_2 /\chi_2^{SM}) < 0.01$, the constraint can be much looser
(see the middle band in Fig.~\ref{fig:GFit_BM3_fig}). Hereafter, the ceiling boundaries of the middle and lower bands will be referred as $1\%$EWPT and $0\%$EWPT respectively.

\section{The LHC and the ILC}\label{sec:LHCILC}

Here we calculate the $Z_s$ Drell-Yan processes at the LHC given the constraints on its couplings obtained above. One needs to fold in the parton distributions of $u\,(d)$ and $\bar{u}\,(\bar{d})$ inside the proton. This involves QCD corrections to the parton model, and are extensively studied with relatively good theoretical control~\cite{Leike:1997cw}. It is also very well tested for the
$Z$ production at the Tevatron. Assume narrow width approximation,
\beq
\frac{q^4}{|q^2-M_{Z_s}^2+i M_{Z_s} \Gamma_{Z_s}|^2}\ra
\delta(q^2-M_{Z_s}^2)\frac{\pi M_{Z_s}^4}{M_{Z_s}\Gamma_{Z_s}}\;,
\eeq
the expected number of observed events, say by reconstructing from the $\mu^+\mu^- X$ final states, is
\beqa
N_{Z_s}= L \sigma_T(pp\ra Z_s X \ra \mu^+\mu^- X )\nonr\\
\simeq  \frac{L}{s} C_{Z_s} C \exp\left[-A \frac{M_{Z_s}}{\sqrt{s}}\right] \,,
\eeqa
where
\beqa
C_{Z_s}(Z_s\ra \mu^+\mu^-)= \frac{4\pi^2}{3}\frac{\Gamma(Z_s)}{M_{Z_s}}
Br(Z_s\ra \mu^+\mu^-)\nonr\\
\times\left[ Br(Z_s\ra u\bar{u}) + \frac{1}{C_{ud}} Br(Z_s\ra d\bar{d})\right] \,,
\eeqa
$L$ is the luminosity, $C_{ud}$ is the ratio of u(d)-parton distribution inside the proton. For the $pp$ hadron collider, $C_{ud}(pp)\sim 2$, $A(pp)\sim 32$, and $C(pp)\sim 600$ for a very wide range of $s$~\cite{Leike:1997cw}.

\begin{figure}[htbp]
\centering
\includegraphics[width=0.4\textwidth]{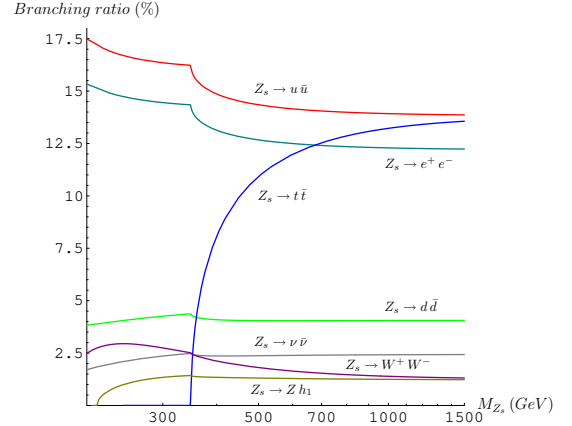}
\caption{Branching ratio for the $Z_s$ decays as functions of $M_{Z_s}$. The curves shown
here are generated with $M_{h_1^0}=120$ GeV and $c_\alpha$, the Higgs mixing angle as defined in
Eq.~(\ref{Eq:HiggsMix}), set to one.}
\label{fig:ZpBr_fig}
\end{figure}

In order to select a signal, it is essential to know the branching ratios of $Z_s$. The calculation for $Z_s$ decays into fermion pair is straightforward. For the SM fermions, they are (again setting
$s_W^2=0.2311$):
\beqa
\Gamma(Z_s\ra u\bar{u})={1 \over 24\pi}{g_2^2\over
c_W^2}M_{Z_s}[0.32739 s_\epsilon^2(1-s_\eta^2)\nonr\\
-0.129957 s_\eta c_\eta s_\epsilon +0.43022 s_\eta^2]\,,\\
\Gamma(Z_s\ra d\bar{d})={1 \over 24\pi}{g_2^2\over c_W^2}M_{Z_s}
[0.09629 s_\epsilon^2(1-s_\eta^2)\nonr\\
-0.277396 s_\eta c_\eta s_\epsilon +0.55451 s_\eta^2]\,,\\
\Gamma(Z_s\ra e\bar{e})={1 \over 24\pi}{g_2^2\over c_W^2}M_{Z_s}
[0.28888 s_\epsilon^2(1-s_\eta^2)\nonr\\
-0.09293 s_\eta c_\eta s_\epsilon +0.12571 s_\eta^2]\,,\\
\Gamma(Z_s\ra \nu \bar{\nu})= {1 \over 24\pi}{g_2^2\over c_W^2}M_{Z_s}
[0.05777 s_\epsilon^2(1-s_\eta^2)\nonr\\
-0.240364 s_\eta c_\eta s_\epsilon + 0.25 s_\eta^2]\,.
\eeqa
After crossing the $t\bar{t}$ threshold, $Z_s$ can decay into a pair of t-quarks. We find
\beq
{\Gamma(Z_s\ra t\bar{t})\over\Gamma(Z_s\ra u\bar{u})}
= \sqrt{1-4\beta} [1 +0.41176\beta +0.7979 \beta^2 ]
\eeq
where $\beta=(M_t/ M_{Z_s})^2$ and the $\beta^2$ term comes from the expansion of $s_\eta/s_\epsilon$.
The decay into a $W^+W^-$ pair has width
\beqa
\Gamma(Z_s\ra W^+W^-)={g_2^2 c_W^2 s_\eta^2 \over 192\pi}{M_{Z_s}^5 \over M_W^4}
(1-4y)^{3/2}\nonr\\
\times(1+20y+12y^2)\,,\; y={M_W^2\over M_{Z_s}^2} \,.
\eeqa
If $M_{Z_s}$ is heavier than $(M_Z+M_{h_1^0})$, and the kinematics are favorable, there is a new decay channel opening up,
\beqa
\Gamma(Z_s\ra Z h_1^0) &\simeq&
{g_2^2 c_\alpha^2  \over 192\pi c_W^2}
(c_\eta +s_Ws_\epsilon s_\eta)^2(s_\eta-s_W s_\epsilon c_\eta)^2\nonr\\
&&\times M_{Z_s}\sqrt{1-4z}(1+8 z )\,,\; z={M_Z^2\over M_{Z_s}^2} \,,
\eeqa
where we have ignored the mass difference between $h^0_1$ and $Z$ to simplify the expression. The branching ratio of shadow Z as a function of its mass is displayed in Fig.~\ref{fig:ZpBr_fig}.
In generating the figure we have treated $M_{Z_s}$ and $s_\epsilon$ as independent parameters and used a very small value of $s_\epsilon=10^{-3}$. This is consistent with the bound of
Fig.~\ref{fig:GFit_BM3_fig}.

In the large $M_{Z_s}$ limit, say $> 1.0$ TeV, the $s_\epsilon^2$-terms will dominate and
we obtain a very simple expression for the decay width:
\beq
\Gamma_{Z_s} \simeq 2.37 { g_2^2 M_{Z_s} s_\epsilon^2 \over 24\pi c_W^2}
= 0.1742\left( {M_{Z_s} \over 1\, \mbox{TeV}}\right)\left( {s_\epsilon^2 \over 0.01}\right)
\mbox{GeV} \,.
\eeq
We see that for such a heavy $Z_s$ its width is indeed very narrow, and the various branching ratios are approximately given by
\beqa
B_u= B_c=B_t &\simeq&  13.81\%\,, \nonr\\
B_d=B_s=B_b &\simeq&  4.06\% \,, \nonr\\
B_e=B_\mu=B_\tau &\simeq&  12.19\%\,, \nonr\\
B_{\nu_e}=B_{\nu_\mu}=B_{\nu_\tau} &\simeq& 2.44\%\,, \nonr\\
B_{W^+W^-} &\simeq& 1.219\%\,, \nonr\\
B_{hZ}&\simeq& 1.219 c_\alpha^2\%\,.
\eeqa
For SM fermions, the branching ratio is roughly proportional to $(Y_L^2+Y_R^2)$. It's interesting to note that $Z_s$ prefers to decay into $u$-type quarks and charged leptons than other SM fermions.
This is very different from the SM $Z$ decay. If the ILC is available, and the $Z_s$ is found, this unique prediction may be tested.

At the LHC with the center of mass energy $\sqrt{s}=14$~TeV and using the bench mark luminosity of
$L=100\,\textrm{fb}^{-1}$,
we calculated the expected number of events of $Z_s$ into different decay modes by
simply folding in the branching ratio we obtained earlier and taking into account of the phase space factors.

\begin{figure}[htbp]
\centering
\includegraphics[width=0.48\textwidth]{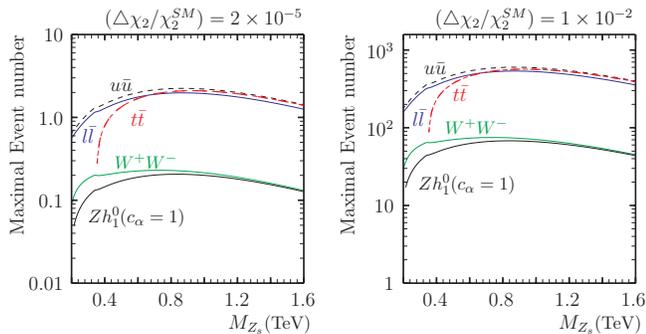}
\caption{The maximal expected number of $Z_s$ events at the LHC for integrated luminosity of
$100 fb^{-1}$ For a fixed $M_{Z_s}$, we have used the largest allowed $s_\epsilon$ comes from the global fit studied in previous section. The left and right panes are for $0\%$EWPT and $0\%$EWPT
respectively.}
\label{fig:LHCLEP}
\end{figure}

More importantly, one has to carefully take into account the maximally allowed $s_\epsilon$
obtained from the global fit of low energy precision measurements as given previously. The
expected number of events depend on how restricted we are in taking the EWPTs. Fig.~\ref{fig:LHCLEP} shows the sensitivity to the $\tri \chi_2/ \chi_2^{SM}$
which can lead to two orders of magnitude difference in the signature. Notice the dipping of the
signals for smaller $M_{Z_s}$. This is due to the much smaller values of $s_\epsilon$ allowed for these relatively light $Z_s$.

One way of distinguishing between different extra $Z$ models will be measuring
 the branching ratios into different fermion species. The $Z_s$ has the feature
 that it has a relatively large branching ratio into charged leptons and the t-quarks.
 For sufficiently heavy $Z_s$ the two branching ratios are almost equal.
 Whether this can be used as a diagnostic tool at the LHC depends on the t-jets and c-jets efficiencies.

\begin{figure}[htbp]
\centering
\includegraphics[width=0.35\textwidth]{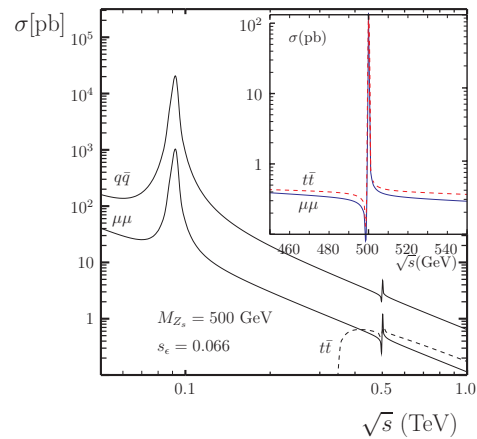}
\caption{The cross section for $e^+e^-\ra f\bar{f}$ with a $500$ GeV $Z_s$ and $s_\epsilon=0.066$.
(In the main frame, the spike tips have been chopped to avoid overlapping among curves.)}
\label{fig:ILCcs}
\end{figure}

The success of LEP has demonstrated that $e^+e^-$ colliders are powerful machines
for studying neutral gauge bosons. Indeed the search of extra $Z$ bosons has been
conducted at LEP and the results can be found in reference~\cite{LEPZ}.
Looking forward to the ILC, the center of mass energy of the collider is lower than
that of the LHC with  $\sqrt s =1$ TeV. We can still expect to see extra $Z$'s of mass
below 1 TeV to be produced. On the other hand, facilities designed to have higher
luminosity and a benchmark integrated luminosity of $500\,\textrm{fb}^{-1}$ can be
expected.
Furthermore, the underlying processes involve much less QCD uncertainties than in hadronic machines, making it a cleaner environment for detecting the extra $Z$ bosons.
Thus, we can anticipate the branching ratios to be accurately measured. Moreover, the $Z_s$ will be too narrow for the total width to be measured. However, spikes will be seen at the mass where LHC ``discovered'' the new state. In Fig.~\ref{fig:ILCcs} we display the result of such a hypothetical occurrence of a $Z_S$ of mass  $M_{Z_s}=500$~GeV and $s_\epsilon=0.066$ corresponding to
the maximal allowed value from the $1\%$EWPT fit. The familiar SM $Z$ boson resonance peaks sit on the left hand side. A new spike appears at $M_{Z_s}$. We magnify the event shape around
$\sqrt s = 0.5$~TeV and we see the characteristic dip at the left base of the peak corresponding to an extra $Z$. This dip is due to the negative contribution from $\gamma-Z_s$ and $Z$-$Z_s$ interference. Although the resonance factor $K_{Z_s Z_s}$ dominates over $K_{Z_s\gamma}$, $K_{Z_sZ}$ around $M_{Z_s}$, the $K_{Z_s Z_s}$ gets an extra $s_\epsilon^2$ suppression in couplings compared to $K_{Z_s\gamma}$ and $K_{Z_sZ}$ which makes the dip visible.

\begin{figure}[htbp]
\centering
\includegraphics[width=0.35\textwidth]{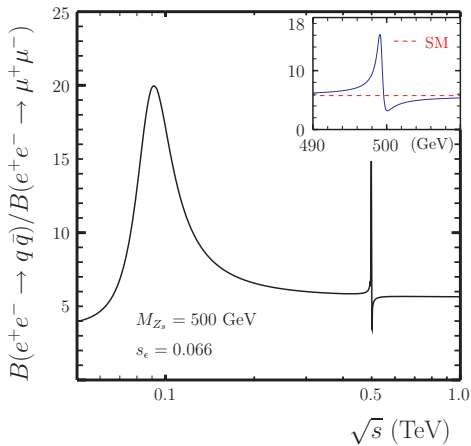}
\caption{The ratio of $e^+e^-\ra \sum_{q \neq t} q\bar{q}$ over $e^+e^-\ra \mu^+\mu^-$ with
a $500$ GeV $Z_s$ and $s_\epsilon=0.066$.}
\label{fig:ILCratio}
\end{figure}

\begin{figure}[htbp]
\centering
\includegraphics[width=0.4\textwidth]{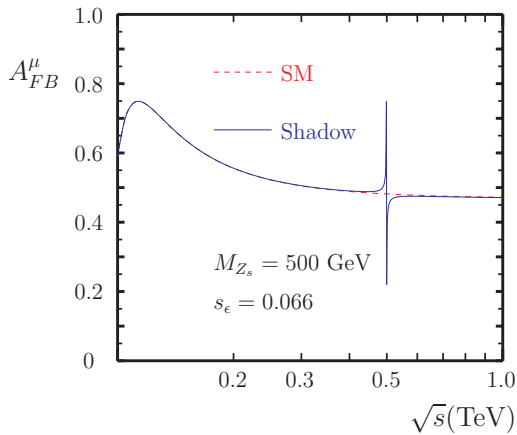}
\caption{The forward-backward asymmetry of muon v.s. $\sqrt{s}$ with
a $500$ GeV $Z_s$ and $s_\epsilon=0.066$.}
\label{fig:ILCAFB}
\end{figure}

For $0\%$EWPT fit, not shown, the spike is not as pronounced and the width is thinner, thus its studies at the ILC will be more challenging.
Similarly, the ratio
\begin{equation}
R^{had}=\frac{Br(e^+e^-\ra q\bar{q})}{Br(e^+e^-\ra\mu^+\mu^-)} \,,
\end{equation}
where $q$ sums over all quarks except the top, has a pronounced spike for $1\%$EWPT fit. This is shown in Fig.~\ref{fig:ILCratio}. The inlay magnifies the region around $Z_s$ and illustrates the expected interference pattern of two spin-1 particles is clearly discernible. The unique event shape is characteristic of this model which may be used to discriminate it from other extra $Z$ models.
Finally, the forward-backward asymmetry of muon v.s. $\sqrt{s}$ is shown in Fig.~\ref{fig:ILCAFB}.

\section{Conclusions}\label{sec:Concl}

We have studied in detail a simple model with an extra neutral $U(1)$ boson, dubbed shadow $Z$.
The shadow $Z$ mixes with the SM $U(1)_Y$ gauge boson kinetically which is parameterized by
$s_\epsilon$. The Higgs sector also contains a scalar field that interacts with the SM Higgs
field. There is no direct coupling between the shadow $Z$ and the SM fermions. This simple model
can easily be embedded in more elaborate model. Since our motivation is purely phenomenological
we leave this aspect to a future study. Instead we embark on a detail analysis of the EWPT constraints and other low energy precision measurements on $M_{Z_s}$ and $s_\epsilon$. It is well known that $s_\epsilon$ can vary greatly from model to model. We found that the data constrain it to be very small for a wide range of extra $Z$ boson masses; see Eq.~(\ref{M3limit}). We conclude that the data favors those models in which $s_\epsilon$ is radiatively generated.

Not surprisingly the production of $Z_s$ at the LHC and the ILC depends crucially on $s_\epsilon$.
In order to ascertain whether the signals are observable  we define a figure of merit measure given
by $\tri\chi_2$ (see Eq.~(\ref{delchi2})) which we found to be positive. We conclude that the shadow $Z$ does not give a better global fit to the EWPTs. However, we use $\tri \chi_2$ to quantify the data tolerance to $Z_s$. We found that $(\tri \chi_2 /\chi_2^{SM})\sim 0.01 $ will lead to an observable production of $Z_s$ via the Drell-Yan process at the LHC.

To distinguish the shadow $Z$ from other extra $Z$ models (see~\cite{PhemZp}) one has to do as many branching ratio measurements as possible. The shadow $Z$ has almost equal branching ratios into u-type quarks and charged leptons. It also has a decay channel into the SM like $Z$ and Higgs boson although it is only 1.3\%. Similarly for the decay into $W^+W^-$ pairs. For this we find that the ILC will be invaluable for pinning down the nature of the extra $Z$ boson.

\acknowledgments{
The research of W.F.C. is supported by the Academia Sinica Postdoctoral Fellowship, Taiwan.
W.F.C. would like to thank the TRIUMF theory group for their kind hospitality where part of this
work is done. J.N.N would like to thank Prof. R. Casalbuoni for providing a warm and stimulating
environment at the GGI, Florence, where the early part of this work was done. He also gratefully
acknowledge the efforts of Prof. B. Vachon of keeping him informed of the experimental situation
in extra $Z$ searches. The authors would like to thank Prof. B. Holdom for useful comments.
This research is partially supported by the Natural Science and Engineering Council of Canada.
}

\section*{Note Added}

After the completion of this paper our attention was drawn to an
earlier work that had looked at similar $Z^\prime$
models~\cite{Appelquist:2002mw}, and also the Stueckelberg
$Z^\prime$ model which has almost identical collider
signatures~\cite{KorsNath} and a similar electroweak
fit~\cite{Feldman:2006ce}. We have checked that our results agree
where they overlap. A variant of the model has also been used in a
recent leptogenesis study~\cite{Cerdeno:2006ha}.


\end{document}